\documentclass[a4paper,conference]{IEEEtran}
\IEEEoverridecommandlockouts

\usepackage{cite}
\usepackage{amsmath,amssymb,amsfonts}
\usepackage{algorithmic}
\usepackage{graphicx}
\usepackage{listings}
\usepackage{textcomp}
\usepackage{xcolor}
\usepackage{balance}
\usepackage{booktabs}
\usepackage{tikz}
\usepackage{pgfplots}
\usepackage{tcolorbox}
\usepackage{adjustbox}
\usepackage{tabularx}
\usepackage{threeparttable}
\usepackage{multirow}
\usepackage{multicol}
\usepackage{url}
\usepackage{hyperref}
\usepackage{colortbl}
\usepackage[capitalize,nameinlink,noabbrev]{cleveref}
\crefname{figure}{Fig.}{Figs.}

\def\BibTeX{{\rm B\kern-.05em{\sc i\kern-.025em b}\kern-.08em
    T\kern-.1667em\lower.7ex\hbox{E}\kern-.125emX}}
\begin{document}

\newcommand{\RqOne}{RQ$_1$: How trustworthy are the LLM-generated security advisories, given a known CVE-ID as input?}

\newcommand{\RqTwo}{RQ$_2$: Do LLMs have the capability to detect fake CVE-IDs?}

\newcommand{\RqThree}{RQ$_3$: Can LLMs accurately and consistently produce CVE-ID from a provided advisory description?}

\title{Using LLMs for Security Advisory Investigations: How Far Are We?
}



\makeatletter
\author{
\begin{@IEEEauthorhalign}
\IEEEauthorblockN{\hspace*{-0.2cm}1\textsuperscript{st} Bayu Fedra Abdullah}
\IEEEauthorblockA{\hspace*{-0.2cm}\textit{Informatics Engineering} \\
\hspace*{-0.2cm}\textit{Universitas Muhammadiyah Surakarta}\\
\hspace*{-0.2cm}Surakarta, Indonesia \\
\hspace*{-0.2cm}L200200115@student.ums.ac.id}
\\ [1.5ex]
\IEEEauthorblockN{\hspace*{-0.2cm}4\textsuperscript{th} Raula Gaikovina Kula}
\IEEEauthorblockA{\hspace*{-0.2cm}\textit{Graduate School of IST} \\
\hspace*{-0.2cm}\textit{University of Osaka}\\
\hspace*{-0.2cm}Osaka, Japan \\
\hspace*{-0.2cm}raula-k@ist.osaka-u.ac.jp}
\and
\IEEEauthorblockN{2\textsuperscript{nd} Yusuf Sulistyo Nugroho}
\IEEEauthorblockA{\textit{Informatics Engineering} \\
\textit{Universitas Muhammadiyah Surakarta}\\
Surakarta, Indonesia \\
yusuf.nugroho@ums.ac.id}
\\ [1.5ex]
\IEEEauthorblockN{5\textsuperscript{th} Kazumasa Shimari}
\IEEEauthorblockA{\textit{Information Science} \\
\textit{Nara Institute of Science and Technology}\\
Nara, Japan \\
k.shimari@is.naist.jp}
\and
\IEEEauthorblockN{\hspace*{0.2cm}3\textsuperscript{rd} Brittany Reid}
\IEEEauthorblockA{\hspace*{0.2cm}\textit{Information Science} \\
\textit{\hspace*{0.2cm}Nara Institute of Science and Technology}\\
\hspace*{0.2cm}Nara, Japan \\
\hspace*{0.2cm}brittany.reid@naist.ac.jp}
\\ [1.5ex]
\IEEEauthorblockN{\hspace*{0.2cm}6\textsuperscript{th} Kenichi Matsumoto}
\IEEEauthorblockA{\hspace*{0.2cm}\textit{Information Science} \\
\hspace*{0.2cm}\textit{Nara Institute of Science and Technology}\\
\hspace*{0.2cm}Nara, Japan \\
\hspace*{0.2cm}matumoto@is.naist.jp}
\end{@IEEEauthorhalign}
}
\makeatother

\maketitle

\begin{abstract}
Large Language Models (LLMs) are increasingly used in software security, but their trustworthiness in generating accurate vulnerability advisories remains uncertain. This study investigates the ability of ChatGPT to (1) generate plausible security advisories from CVE-IDs, (2) differentiate real from fake CVE-IDs, and (3) extract CVE-IDs from advisory descriptions. Using a curated dataset of 100 real and 100 fake CVE-IDs, we manually analyzed the credibility and consistency of the model’s outputs.
The results show that ChatGPT generated plausible security advisories for 96\% of given input real CVE-IDs and 97\% of given input fake CVE-IDs, demonstrating a limitation in differentiating between real and fake IDs. Furthermore, when these generated advisories were reintroduced to ChatGPT to identify their original CVE-ID, the model produced a fake CVE-ID in 6\% of cases from real advisories.
These findings highlight both the strengths and limitations of ChatGPT in cybersecurity applications. 
While the model demonstrates potential for automating advisory generation, its inability to reliably authenticate CVE-IDs or maintain consistency upon re-evaluation underscores the risks associated with its deployment in critical security tasks.
Our study emphasizes the importance of using LLMs with caution in cybersecurity workflows and suggests the need for further improvements in their design to improve reliability and applicability in security advisory generation.
\end{abstract}

\begin{IEEEkeywords}
advisory, chatgpt, cve id, security, vulnerability
\end{IEEEkeywords}

\section{Introduction}
\label{sec:intro}

In recent years, software vulnerabilities have become the main focus for software developers, especially due to the increase in cybercrimes~\cite{sanchez2019software}. Developers may attempt to mitigate vulnerabilities~\cite{nina2021systematic}, but hackers continue exploiting weaknesses to attack users and steal sensitive data~\cite{mishra2021impact}. To aid the secure development of software, the Common Vulnerabilities and Exposures (CVE) system maintains a database\footnote{\url{https://www.cve.org/}} of publicly disclosed vulnerabilities in software packages, labeling each with a unique CVE-ID and providing advisory information. CVEs are a common way of disseminating security information within the software community, and thus recognizing and identifying them is important to the security of software. 


With the rise of generative AI, developers increasingly turn to Large Language Models (LLMs) for assistance in various domains, such as health~\cite{daungsupawong2024innovative}, education~\cite{adiguzel2023revolutionizing}, programming~\cite{liu2024refining}, including software security~\cite{gupta2023chatgpt}. 
Despite their growing popularity, concerns remain regarding the accuracy and reliability of outputs~\cite{rhiannon2024why}.  
In cybersecurity, some practitioners use tools like ChatGPT to retrieve CVE-IDs or generate advisories, though LLMs' trustworthiness in this context is uncertain. While LLMs show promise in vulnerability detection~\cite{noever2023can} and automated security bug repair~\cite{pearce2023examining}, their tendency to hallucinate information~\cite{ji2022survey} raises risks.
Relying on AI-generated advisories without verification may create false security assumptions or even delay critical mitigation efforts. Although LLMs are not designed as database query systems, our study reflects real-world misuse scenarios where users treat them as trusted security sources.

This study investigates the capabilities of LLMs in generating accurate and reliable security advisories, identifying inconsistencies in vulnerability data, and detecting fake CVE-IDs. 
Prior research has examined LLMs in vulnerability detection and code security analysis. However, limited work has explored their reliability in security advisory generation. To address this gap, we formulate the following research questions:

\begin{itemize}
    \item \emph{\RqOne}
    \item \emph{\RqTwo}
    \item \emph{\RqThree} 
\end{itemize}


To answer these questions, we constructed a dataset comprising 100 real CVE-IDs and their corresponding security advisories from the GitHub Security Advisory (GHSA) database,\footnote{\url{https://github.com/advisories}} capturing key metadata such as CVE-ID, title, severity, and affected products. To evaluate the LLM's ability to detect fabricated vulnerabilities, we generated an additional 100 fake CVE-IDs using a Python script and verified their absence in public vulnerability databases. Both real and fake CVE-IDs were provided to ChatGPT, and the generated advisories were analyzed for accuracy, consistency, and reliability.

The findings highlight both the potential and the risks of utilizing LLMs for security advisory generation. While LLMs show impressive generative capabilities, their tendency to produce misinformation raises concerns about their deployment in critical cybersecurity tasks. Our study highlights the importance of validating AI-generated security advisories before adoption and serves as a foundation for future research aimed at improving LLM reliability in cybersecurity applications. 

\section{Related Work}
\label{sec:relwork}

\subsection{LLMs for Software Engineering}
Since the first release of the LLM, numerous studies have been conducted on its implementation in software engineering.
Studies reveal that LLMs or generative AI tools like Google Bard (now called Gemini), ChatGPT, or CoPilot have proven that they can improve productivity in software engineering~\cite{brie2023evaluating, ebert2023generative}.
The use of LLMs has also benefited many software engineering tasks. 
For example, LLMs can accelerate the development cycles and reduce the time spent on repetitive coding tasks~\cite{khomh2023harnessing}, increasing productivity~\cite{ross2023programmer}, improving software quality and development efficiency~\cite{yu2023llm}, and even the LLMs can help in identifying the software vulnerability-related subtle patterns~\cite{purba2023software}.

Despite the advantages of the use of LLMs in software engineering, it has been unknown how far the LLMs can distinguish between real or fake vulnerabilities and their descriptions.

\subsection{Identifying Software Vulnerabilities}

As one of the weaknesses in computer security, vulnerabilities are often used by an attacker to exploit the system to perform unauthorized actions~\cite{munaiah2019characterizing, piessens2016software}. 
To mitigate this, developers sometimes notify other programmers by citing a specific vulnerability identifier in code comments to indicate that the code may contain vulnerability~\cite{nugroho2022study}.
Other strategies were also proposed in several researches, such as the development of automatic identification of vulnerable versions for CVE~\cite{bao2022vszz}, the implementation of TF-IDF and Doc2Vec for automatically tracing the related CAPEC-IDs from CVE-ID~\cite{kanakogi2021tracing}, and a machine-learning based technique to assign pertinent CWE identifiers to new CVE entries~\cite{aota2021multi}.
These trigger software developers to make efforts and help them in addressing security issues consistently.

In addition, some prior works were conducted by focusing on the vulnerability descriptions. For example, a study proposed a novel method to compose CVE descriptions by extracting some vulnerability aspects from ExploitDB~\cite{sun2021generating}.
Their results indicate that the method can achieve high accuracy in composing CVE descriptions.
Another study augmented the vulnerability description by scrapping third-party references (hyperlinks)~\cite{althebeiti2023enriching}.
The findings have shown potential regarding summary fluency, completeness, correctness, and understanding.
Due to the increasing of AI implementations in the software engineering field, the CVE description can also be potentially written with the help of AI, such as ChatGPT.
Thus, in this paper, we study to what extent an LLM can generate the security advisory description.





\section{Study Setting}

\subsection{Research Questions}
\label{sec:rqs}

To guide our study, we formulated the following three research questions and their motivations.

\subsection*{\textbf{\RqOne}}

\textbf{Motivation:} 
This RQ arises from the increasing use of LLMs like ChatGPT in cybersecurity. Many developers and security practitioners rely on these models to generate security advisories and identify vulnerabilities. However, the accuracy of AI-generated advisories remains uncertain. Since cybersecurity decisions require precise and reliable information, any incorrect or misleading advisory could result in overlooked vulnerabilities or a false sense of security.

In addition, LLMs are known to hallucinate~\cite{ji2022survey}, generating plausible but false information. In the case of security advisories, this could mean fabricated vulnerability details or incorrect mitigation steps. Since non-expert users may not have the expertise to validate AI-generated advisories, it is crucial to assess how trustworthy these outputs are. By investigating the trustworthiness of LLM-generated security advisories, we aim to evaluate whether AI models can serve as a reliable source of cybersecurity information or if their use introduces risks that could undermine software security efforts. 

\begin{figure*}[]
    \centering
    \includegraphics[width=.75\linewidth]{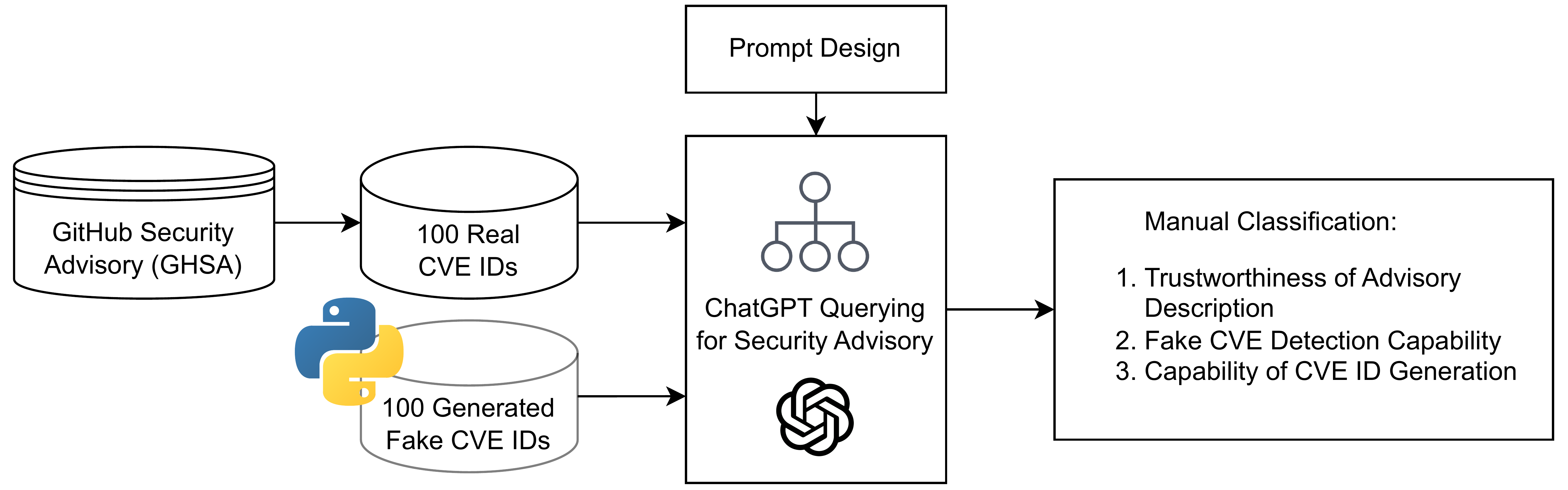}
    \caption{Overview of the research procedures, covering data collection, prompt design, and manual analyses.}
    \label{fig:researchprocedure}
\end{figure*}

\subsection*{\textbf{\RqTwo}}
\textbf{Motivation:}
While databases like CVE and GHSA provide authoritative records of known vulnerabilities, developers and security practitioners may still turn to AI tools like ChatGPT for quick information retrieval. However, the ability of LLMs to differentiate real vulnerabilities from fabricated ones remains unclear. If an LLM fails to detect fake CVE-IDs, it could generate misleading security advisories, leading to confusion or even security misconfigurations. This is particularly concerning because attackers or misinformed users could exploit AI-generated misinformation to spread false vulnerabilities or obscure real threats.

Moreover, LLMs are trained on large amounts of publicly available text but lack direct access to up-to-date vulnerability databases. Without real-time verification mechanisms, they may produce CVE-IDs that sound reasonable but are entirely fake or fail to recognize invalid ones. 
By investigating whether LLMs can effectively detect fake CVE-IDs, this study aims to assess their reliability as a security resource and highlight potential weaknesses in AI-driven vulnerability identification.

\subsection*{\textbf{\RqThree}}
\textbf{Motivation:}
When analyzing vulnerabilities, correctly associating an advisory with its corresponding CVE-ID is essential for tracking and mitigating security risks. If LLMs can accurately generate CVE-IDs from advisory descriptions, they could serve as useful tools for automating parts of the vulnerability management process. 
This would benefit developers, security teams, and organizations by improving efficiency in identifying and responding to threats. 
However, if LLMs generate incorrect CVE-IDs, they risk spreading misinformation and creating confusion within the security community.

This limitation raises concerns about their ability to consistently match advisories with the correct CVE-IDs, especially when dealing with newly disclosed vulnerabilities. If an LLM frequently generates incorrect CVE-IDs, this could undermine trust in the security insights generated by AI. By evaluating the accuracy of LLMs in generating CVE-IDs, this study aims to determine their usefulness in cybersecurity workflows and highlight areas where improvements are necessary to ensure their reliability as an information source.

\subsection{Data Collection}
\label{sec:data_collection}

As illustrated in~\cref{fig:researchprocedure}, the entire process of this study includes data collection, fake CVE IDs generation, prompt design, ChatGPT querying, and manual classification.

The dataset used in this study, as presented in~\autoref{tab:dataset}, was built through the following 2 main steps:

\begin{table}[]
    \centering
    \caption{Dataset Overview}
    \begin{tabular}{|l|c|}
        \hline
        \textbf{Types of CVE ID} & \textbf{\# CVE IDs}  \\
        \hline
        Real CVE-IDs &  100 \\
        Generated CVE-IDs &  100    \\
        \hline
        \textbf{Total} & 200    \\
        \hline
    \end{tabular}
    \label{tab:dataset}
\end{table}

\subsubsection{Extracting Security Advisory info from GHSA}
To build our dataset, we initially downloaded all GitHub Security Advisory (GHSA) data. 
We then randomly selected 100 real CVE-IDs and their advisory descriptions by specifying the attributes including CVE-ID, title, affected product, severity, CWE-ID, published date, and details, as exemplified in~\cref{fig:GHSAexample}.

\begin{figure}[]
    \centering
    \begin{minipage}{.96\columnwidth}
    \footnotesize
    \begin{lstlisting}[basicstyle=\ttfamily, frame=single]
cve_id: CVE-2016-3722
title: Incorrect Authorization in Jenkins Core
affected: Maven org.jenkins-ci.main:jenkins-core
severity: MODERATE
cwe_id: CWE-863
published: 2022-05-14
details: Jenkins before 2.3 and LTS before 1.651.2
allow remote authenticated users with multiple
accounts to cause a denial of service (unable to
login) by editing the "full name".
    \end{lstlisting}
    \end{minipage}
    \caption{Example of extracted security advisory from GHSA.}
    \label{fig:GHSAexample}
\end{figure}

\subsubsection{Generating Fake CVE-IDs}
To understand how far we can rely on LLMs in generating security advisories, we generated 100 random fake CVE-IDs automatically using Python code by following the format \texttt{CVE-YYYY-NNNN(N)}. To ensure that the generated CVE-IDs were fake, we cross-checked the identifiers on the NVD Database.\footnote{https://nvd.nist.gov/} If the CVE IDs exist in the database, they were ignored. Otherwise, they were kept for further analysis of fake security advisories. Some generated fake CVE-IDs are shown in~\cref{fig:fakeCVEIDs}.

\begin{figure}[]
    \centering
    \begin{minipage}{.96\columnwidth}
    \scriptsize
    \begin{lstlisting}[basicstyle=\ttfamily, frame=single]
cve-2021-6365
cve-2016-13580
cve-2018-8981
cve-2022-13790
cve-2022-6453
    \end{lstlisting}
    \end{minipage}
    \caption{Example of fake CVE-IDs.}
    \label{fig:fakeCVEIDs}
\end{figure}

\subsection{Prompt Design}
Prompts using CVE-IDs replicate actual developer behavior observed on platforms like Stack Overflow and Reddit. For instance, a Reddit discussion\footnote{\url{https://www.reddit.com/r/artificial/comments/1345ay8/chatgpt_leaks_reserved_cve_details_should_we_be/}} illustrates how users query ChatGPT regarding known and even reserved CVE details, highlighting its usage in real-world security contexts.
Prompts were entered manually into ChatGPT using a structured template, as shown in~\cref{fig:prompts1}. In this case, no files or fine-tuning were used. 
We also observed that ChatGPT’s responses were highly sensitive to the phrasing of prompts; for example, including explicit instructions such as ``mention the affected product explicitly'' produced more detailed information, while omitting such cues often resulted in vague or generic outputs. These findings emphasize the importance of careful prompt design, which we plan to explore further in future work.

\begin{figure}[]
    \centering
    \begin{minipage}{.96\columnwidth}
    \scriptsize
    \begin{lstlisting}[basicstyle=\ttfamily, frame=single]
You will provide information about the given CVE ID. 
Your output should follow these rules:
- Provide the title of the CVE in the "Title" label
- Mention the affected product explicitly and its version 
  in the "Affected" label
- Define the severity level in the "Severity" label
- Provide the CWE ID in the "CWE-ID" label
- Give the "Published date" in 'YYYY-MM-DD' format in the 
  "Published" label
- Write the description of the given CVE ID in the 
  "Description" label
- Return label only without other text
    \end{lstlisting}
    \end{minipage}
    \caption{Example of input to LLM, asking for vulnerability advisory descriptions.} 
    \label{fig:prompts1}
\end{figure}

\subsection{Manual Classification}
The manual classification in this study was conducted by a single evaluator following a structured rubric. For trustworthiness, advisories were labeled as `Reliable' if they appeared internally consistent and plausible, and `Unreliable' if they had inconsistencies, such as date mismatches or implausible information. For similarity, the evaluator used a five-level scale ranging from `Totally Different' to `Similar,' based on both lexical overlap and semantic coherence.
In detail, we describe them in each RQ in Section~\ref{sec:result}.

\subsection{Appendix}
To facilitate the reproducibility of this work, we made our replication package publicly available on~\url{https://github.com/bayufedra/Research-NAIST-SE-2024}.

    







\section{Results and Discussions}
\label{sec:result}

In this section, we describe the findings of each research question and discuss them comprehensively.

\subsection{\RqOne}

\begin{table}[]
    \centering
    \caption{The Trustworthiness of the LLM-Generated Advisory Descriptions}
    \begin{tabularx}{\linewidth}{|l|X|}
        \hline
        \textbf{Category}   & \textbf{Descriptions} \\
        \hline
        Reliable & If the generated security advisory is consistent and looks credible.\\
        Unreliable &  If the generated security advisory exhibits inconsistencies and seems unreliable. \\
        \hline  
    \end{tabularx}
    \label{tab:trustworthinessCategory}
\end{table}

To address this question, we asked ChatGPT to generate security advisory details according to the given CVE-ID. By using the designed prompt in~\cref{fig:prompts1}, we input both real and fake CVE-IDs. All LLM outputs were then manually analyzed to evaluate the accuracy of the generated descriptions.

As described in~\autoref{tab:trustworthinessCategory}, the LLM-generated outputs were classified into two categories: \textit{Reliable} and \textit{Unreliable}. We labeled the output \textit{Reliable} if there is a possibility for someone to believe that the security advisory is real. Otherwise, it was labeled \textit{Unreliable}, including inconsistencies. For example, the published year of the advisory is different from the year indicated in the CVE-ID code.

The result, as presented in~\autoref{tab:trustworthiness}, shows that ChatGPT mostly generates \textit{reliable} advisories, accounting for 96\% and 97\% for real and fake CVE-IDs, respectively. 
This indicates that ChatGPT is remarkably good at generating reliable output, regardless of whether the input is real or not. Thus, it may not be obvious to developers if the output or input is misleading since ChatGPT does not detect whether the CVE-IDs are fake.

\begin{table}[]
    \centering
    \caption{Trustworthiness of ChatGPT in Generating Security Advisory Descriptions Based on the Given CVE-IDs}
    \label{tab:trustworthiness}
    \begin{tabular}{|l|r|r|}
        \hline
        \multirow{2}{*}{\textbf{Type of CVE-ID}} & \multicolumn{2}{c|}{\textbf{\# CVE-IDs}} \\
        \cline{2-3}
         & \textbf{Reliable} & \textbf{Unreliable}  \\
        \hline
        Real & 96\% & 4\% \\
        Fake & 97\% & 3\%   \\
        \hline
    \end{tabular}
\end{table}

Since the trustworthiness of security advisories generated by ChatGPT does not necessarily reflect the accuracy of the associated CVE-IDs, a deeper analysis was required. In this case, we only focused on the 100 randomly chosen real CVE-IDs. Therefore, we examined the similarity between the advisory descriptions of these real CVE-IDs and the descriptions generated by ChatGPT.
In this analysis, we manually classified the descriptions into five classes based on the similarity, as described in~\autoref{tab:compliance}.

\begin{table}
    \centering
    \caption{Compliance Categories Between LLM-Generated and Original Security Advisory Descriptions}
    \label{tab:compliance}
    \resizebox{\columnwidth}{!}{
    \begin{tabular}{|l|p{6.5cm}|}
        \hline
        \textbf{Category} & \textbf{Descriptions}    \\
        \hline
        Totally Different &  If the generated description is entirely unrelated.\\
        Quite Different &  If the generated output has some connections but most of the information is different.  \\
        Somewhat Different & If the description generated by LLM has significant connections but there are differences in details or structure\\
        Quite Similar & If the output is mostly the same, with only minor or insignificant differences \\
        Similar & If the generated description is exactly the same or fully aligned without any noticeable differences \\
        \hline
    \end{tabular}%
    }
\end{table}





\begin{figure}
    \begin{tikzpicture}
        \begin{axis}[
            xbar,                              
            symbolic y coords={Totally Diff, Quite Diff, Somewhat Diff}, 
            ytick=data,                    
            xmin=0,      
            xlabel={Percentage},       
            bar width=10pt,
            height=3.2cm,
            width=.9\columnwidth,
            enlarge y limits=0.3,              
            nodes near coords={\pgfmathprintnumber\pgfplotspointmeta\%},
            every node near coord/.append style={font=\tiny}, 
            x tick label style={font=\scriptsize},
            y tick label style={font=\scriptsize},
            xlabel style={font=\scriptsize},
            ylabel style={font=\scriptsize}
        ]
        \addplot [draw=none, fill=gray] coordinates {
            (95,Totally Diff) 
            (3,Quite Diff) 
            (2,Somewhat Diff)};
        \end{axis}
    \end{tikzpicture}
    \caption{Similarity distribution of ChatGPT-generated advisories to the original ones.}
    \label{fig:horizontalbarchart}
\end{figure}
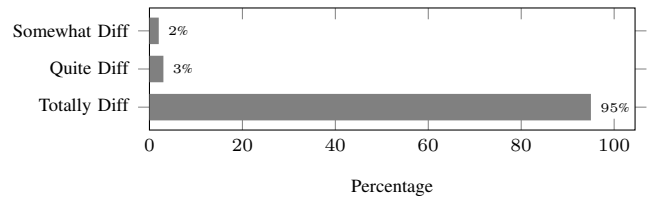


As can be seen in~\cref{fig:horizontalbarchart}, it shows that ChatGPT produces 95\% outputs that totally different from the original advisory descriptions, 3\% are quite different, and 2\% are somewhat different. Notably, none of the generated advisories are classified as ``Quite Similar'' or ``Similar,'' highlighting the inability of the model to accurately replicate original advisory content.

These results suggest that although ChatGPT can generate security advisories, its descriptions are often unreliable and lack fidelity to the original CVE advisories. This poses a significant risk, as practitioners relying on LLM-generated advisories may receive misleading or entirely incorrect security information. The high rate of ``Totally Different'' advisories also underscores the model's tendency to hallucinate content rather than align with verified vulnerability data. Given these findings, it is critical for security professionals to verify AI-generated advisories against trusted sources before relying on them for decision-making in cybersecurity contexts.

\subsection{\RqTwo}


In this RQ, we examined the ability of ChatGPT to detect its own generated fake CVE-IDs. We inputted each fake ID of the CVE into the model and asked it to detect whether the IDs were real or fake.

As described in~\autoref{tab:freqLLMdetectingfakeCVEID}, ChatGPT failed to flag any fake CVE-IDs as invalid. However, this does not imply complete model failure but rather reflects a lack of uncertainty indication in LLM responses.
This result demonstrates that ChatGPT cannot validate CVE-IDs against an authoritative database. 
Users who rely on ChatGPT to verify CVE-IDs may unknowingly accept and propagate fabricated vulnerabilities, which could have serious implications for cybersecurity decision-making.


\begin{table}[]
    \centering
    \caption{Failure Rate of ChatGPT to Detect Fake CVE-IDs}
    \label{tab:freqLLMdetectingfakeCVEID}
    \begin{tabular}{|l|r|}
        \hline
        \textbf{Detection Result} & \textbf{\# Fake CVE-IDs} \\
        \hline
        Detected & 0\% \\
        Not detected & 100\%   \\
        \hline
    \end{tabular}
\end{table}

The inability of ChatGPT to differentiate real CVE-IDs from fake ones highlights a fundamental limitation in its knowledge retrieval and fact-checking processes. This limitation poses a significant risk, for example, bad actors could exploit the inability of ChatGPT to detect fakes by spreading false vulnerability reports, potentially causing unnecessary security issues or misleading developers into believing non-existent threats are real.
These findings emphasize the importance of integrating external validation mechanisms, 
before relying on LLM-generated security advisories.

\subsection{\RqThree}

\begin{figure}[]
    \centering
    \begin{minipage}{.96\columnwidth}
    \scriptsize
    \begin{lstlisting}[basicstyle=\ttfamily, frame=single]
You will show a CVE-ID based on the given Security 
Advisory. Your output must follow these rules:
- Give CVE-ID of the Security Advisory in the "CVE-ID" 
  label.
- Return label only without other text.
    \end{lstlisting}
    \end{minipage}
    \caption{LLM input to ask for CVE-ID from a given advisory.} 
    \label{fig:prompts2}
\end{figure}

In this question, we used both 100 real CVE advisories extracted from GHSA database, 100 generated advisories of these original real CVE-IDs resulting from RQ$_1$, and 100 generated advisories of fake CVE-IDs. Subsequently, we asked ChatGPT whether it could identify the CVE-ID code from the advisory text using the designed prompt, as shown in \cref{fig:prompts2}. 

\begin{table}[]
    \centering
    \caption{Success Rate of ChatGPT in Generating CVE-IDs Based on the Real Advisories}
    \label{tab:freqCVEIDAnswer}
    \begin{tabular}{|l|r|}
        \hline
        \textbf{Type of Correctness} & \textbf{\# CVE-IDs} \\
        \hline
        True CVE ID & 94\% \\
        False CVE ID & 6\%   \\
        \hline
    \end{tabular}
\end{table}

First, we asked ChatGPT to generate the CVE-ID based on 100 real advisories. We considered ``True'' if the generated CVE-ID was fully correct. Otherwise, we marked it as ``False'' even if it partially matched the true CVE-ID (e.g., correct year but wrong number). While this approach is strict, it ensures clarity, and future work may adopt graded scoring to capture partial correctness. As shown in~\autoref{tab:freqCVEIDAnswer}, ChatGPT was incapable of generating 6\% CVE-IDs. 

Next, we asked ChatGPT to identify the CVE-ID based on the advisories generated from the same set of real CVE-IDs. This analysis was conducted to assess its knowledge consistency. We found that, out of 100 advisories, only one had a consistent answer, as shown in~\autoref{tab:LLMconsistencytogenerateCVEID}. This suggests that while ChatGPT is capable of generating realistic-sounding security advisories, it struggles to accurately generate unique and valid CVE-IDs.


\begin{table}[]
    \centering
    \caption{The Consistency of the LLM in Generating CVE-IDs from Its Own Generated Security Advisory}
    \label{tab:LLMconsistencytogenerateCVEID}
    \begin{tabular}{|l|r|}
        \hline
        \textbf{Consistency} & \textbf{\# CVE-IDs} \\
        \hline
        Consistent CVE ID & 1\% \\
        Inconsistent CVE ID & 99\%   \\
        \hline
    \end{tabular}
\end{table}

To add to our understanding of its ability, we further asked ChatGPT with the same scenario as in the previous experiment using the same prompt as designed in~\cref{fig:prompts2}. However, in this case, we utilized the 100 generated security advisories based on fake CVE-IDs yielded from RQ$_1$ as the input.
The results, as shown in~\autoref{tab:freqCVEIDAnswer2}, show that 10\% of the CVE-IDs generated by ChatGPT are detected in the database even though the advisory given is false. This further highlights the limitation of the model in accurately generating unique and valid CVE-IDs, even when provided with fabricated input data.


\begin{table}[]
    \centering
    \caption{Frequency of CVE-ID Existence Detection Based on Fake Advisories}
    \label{tab:freqCVEIDAnswer2}
    \begin{tabular}{|l|r|}
        \hline
        \textbf{Authenticity} & \textbf{\# CVE-IDs} \\
        \hline
        Existing CVE ID & 10\% \\
        Fabricated CVE ID & 90\%   \\
        \hline
    \end{tabular}
\end{table}

\section{Threats to Validity}
One potential threat to validity in this study relates to the temporal nature of LLM capabilities. Our research was conducted between April and June 2024, meaning that all results presented in this paper reflect the state of ChatGPT during that period. Since LLMs continuously evolve through updates and improvements, their performance in generating and verifying security advisories may change over time. 

Another threat concerns the prompt design used to query ChatGPT. The effectiveness of LLM responses can be highly sensitive to the wording, structure, and specificity of prompts. 
Furthermore, our study relied on manual labeling to classify advisory similarities and assess the correctness of generated CVE-IDs. This process might be subjective, as human evaluators may interpret and categorize responses differently. Although we maintained consistency in our labeling approach, variations in judgment among different researchers could introduce bias into the results.

\section{Conclusion}
\label{sec:conclusion}
This study evaluated the capabilities of an LLM, specifically ChatGPT in generating credible security advisories and identifying CVE-IDs. We investigated the performance of ChatGPT in identifying 100 real and 100 fake CVE-IDs to gain valuable insights into its limitations in cybersecurity applications.




Our findings reveal the limitations in the ability of ChatGPT to generate accurate and trustworthy security advisories, detect fake CVE-IDs, and correctly assign CVE-IDs to advisories. Although the model can produce realistic-sounding advisories, it shows that the generated descriptions frequently differ from the original ones. ChatGPT also fails to detect fake CVE-IDs, as it consistently identifies them as real. Furthermore, its ability to generate CVE-IDs from advisory descriptions is inconsistent, with a higher percentage of incorrect outputs. These findings suggest that while LLMs can assist in security-related tasks, they should not be relied upon to create or verify security advisory without human supervision. Future research should explore ways to improve the reliability of LLMs in security contexts, such as integrating external validation mechanisms, refining prompt engineering strategies, and improving dataset quality for training security-related models.


\bibliographystyle{IEEEtran}

\balance

\end{document}